\documentclass[11pt]{article}
\usepackage[top = 2 cm, bottom = 2 cm, left = 2 cm, right = 1.8 cm]{geometry}
\usepackage{authblk, cite, color, amssymb, amsmath}
\usepackage[hypertex, colorlinks = true, linkcolor = blue, citecolor = red]{hyperref}
\usepackage[dvipsnames]{xcolor}

\begin{document}

\title{Fixing cosmological constant on the event horizon}

\author[1,2]{Merab Gogberashvili}
\affil[1]{Javakhishvili State University, 3 Chavchavadze Ave., Tbilisi 0179, Georgia}
\affil[2]{Andronikashvili Institute of Physics, 6 Tamarashvili St., Tbilisi 0177, Georgia}
\maketitle

\begin{abstract}
Standard cosmological equations are written for the Hubble volume, while the real boundary of space-time is the event horizon. Within the unimodular and thermodynamic approaches to gravity, the dark energy term in cosmological equations appears as an integration constant, which we fix at the event horizon and obtain the observed value for the cosmological constant.

\end{abstract}
\vskip 5mm

%%%%%%%%%%%%%%%%%%%%%%%%%%%%%%%%%%%%%%%%%%%%%%%%%%%%%%%%%%%%%%%%%%%%%%%%%%%%%%%

Numerous observations imply that the late universe is in accelerated expansion \cite{SupernovaCosmologyProject:1998vns, SupernovaSearchTeam:1998fmf}. Within the framework of the standard cosmological model, an accelerated expansion can be accounted for by a positive cosmological constant $\Lambda$, which presents in the system of equations for a homogeneous, isotropic and flat universe ($k = 0$),
\begin{eqnarray}
H^2 &=& \frac{8\pi G}{3} \rho + \frac \Lambda 3 ~, \label{FRIEDMANN}\\
\frac {\ddot a}{a} \equiv \dot{H} + H^{2} &=& - \frac{4\pi G}{3} (\rho + 3p) + \frac \Lambda 3 ~. \label{accelerate}
\end{eqnarray}
Here overdots denote derivatives with respect to the cosmic time, $H \equiv \dot{a}/a$ is the Hubble parameter, $G$ is the gravitational constant, and $\rho$ and $p$ are the mass and pressure densities of cosmological fluids (we use the system of units where $c = \hbar = k_B = 1$).

In (\ref{FRIEDMANN}) and (\ref{accelerate}) the cosmological constant $\Lambda$ is a free parameter and is fixed only from observations, which for the value of dark energy density gives \cite{Planck:2015fie},
\begin{equation} \label{Observ}
\Omega_{\Lambda} = \frac {\Lambda}{3H_0^2} \approx 0.692 ~,
\end{equation}
where $H_0$ denotes the present value of the Hubble parameter. One of the main problems of standard cosmology is that the measured value of $\Lambda$ is much smaller than theoretical estimations obtained from the standard quantum field theory. In order to resolve this discrepancy, various models have been proposed (see the reviews \cite{Padmanabhan:2002ji, Bamba:2012cp}).

It is known that, using the Friedmann equation (\ref{FRIEDMANN}), the acceleration equation (\ref{accelerate}) can be expressed without $\Lambda$,
\begin{equation} \label{dotH}
\dot{H} = - 4\pi G (\rho + p) ~.
\end{equation}
Also, combining (\ref{FRIEDMANN}) and (\ref{accelerate}), one can obtain the matter energy-momentum conservation equation,
\begin{equation} \label{dotRho}
\dot{\rho} = - 3 H (\rho + p) ~,
\end{equation}
where the cosmological constant $\Lambda$ does not appear as well. If instead of (\ref{FRIEDMANN}) and (\ref{accelerate}), one will choose (\ref{dotH}) and (\ref{dotRho}) as the independent system of cosmological equations, $\Lambda$ obtains the role of integration constant. Indeed, excluding $(\rho + p)$ from (\ref{dotH}) and (\ref{dotRho}),
\begin{equation}
\frac {4\pi G}{3} \dot{\rho} = H\dot{H} ~,
\end{equation}
and integrating this relation over the time, we obtain the Friedmann equation (\ref{FRIEDMANN}),
\begin{equation} \label{balance}
H^2 = \frac {8\pi G}{3} \rho + C ~,
\end{equation}
but with an integration constant $C$ instead of $\Lambda/3$. So, in a cosmological models where the system (\ref{dotH}) -- (\ref{dotRho}) is primary, and the Friedmann equation (\ref{FRIEDMANN}) represents its first integral, the value of $\Lambda$ can be obtained from boundary conditions of the model. The examples of scenarios where $\Lambda$ arises as an integration constant are unimodular relativity \cite{Ng:1990xz, Finkelstein:2000pg, Padilla:2014yea, Carballo-Rubio:2022ofy} and thermodynamic approach to gravity \cite{Padmanabhan:2012ik, Padmanabhan:2012gx, Komatsu:2018meb, Gogberashvili:2018jkg, Gogberashvili:2014ora, Jacobson:1995ab, Padmanabhan:2009vy}.

Key ingredient in thermodynamic model of gravity is entropy, which allows us to study different aspects of various physical systems using a similar mathematical framework (see the recent review \cite{Ribeiro-2021}). Most thermodynamic cosmological scenarios are based on the holographic principle (see the review \cite{Bousso:2002ju}) and for an associated entropy of a volume usually is used the Bekenstein--Hawking formula for black holes \cite{Bekenstein:1973ur, Hawking},
\begin{equation} \label{S_BH}
S_{\rm{BH}} = \frac{A}{4G} = \frac{\pi R^2}{G} ~,
\end{equation}
where $R$ denotes the radius and $A = 4\pi R^2$ is the surface area. In thermodynamic cosmology, to the Hubble sphere of radius
\begin{equation}
R_H = \frac 1H ~,
\end{equation}
having the surface and volume
\begin{equation} \label{V_H}
A_H = \frac {4\pi}{H^2}~, \qquad V_H = \frac {4\pi}{3H^3} ~,
\end{equation}
can be associated the temperature \cite{Hawking:1975vcx},
\begin{equation} \label{T_H}
T_H = \frac {H}{2\pi} ~,
\end{equation}
and the black hole type entropy (\ref{S_BH}),
\begin{equation} \label{S_H}
S_H = \frac {A_H}{4G} = \frac{\pi}{GH^2} > 0 ~.
\end{equation}
In thermodynamic approach, the Friedmann equation (\ref{balance}) has the natural interpretation as the balance of gravitational and matter heat densities in the spirit of the first law of thermodynamics \cite{Jacobson:1995ab, Padmanabhan:2009vy}, while the acceleration equation (\ref{dotH}) can be obtained by equating the entropy input in the Hubble volume to the sum of entropy flux (entropy received per unit surface) transferred through the horizon and the entropy supplied by internal sources (entropy generated per unit volume). Indeed, if we neglect the entropy supplied by internal sources, the time derivative of the entropy contained within the Hubble volume, $S_H$, should be equal to the flux of the matter entropy density, $S_m$, through the boundary $A_H$,
\begin{equation} \label{dS=}
\dot S_H = S_m A_H ~.
\end{equation}
Using the classical Gibbs--Duhem relation,
\begin{equation}\label{Gibbs-Duhem}
\rho + p = T_HS_m ~,
\end{equation}
the equation (\ref{dS=}) takes the form:
\begin{equation} \label{dA/dt}
\dot A_H = 4G (\rho + p)\, \frac{A_H}{T_H} ~,
\end{equation}
which, using (\ref{V_H}) and (\ref{T_H}), reduces to the standard acceleration equation (\ref{dotH}). Note that in thermodynamic picture the Universe cannot export the entropy to any external universes and it seems that its total entropy is conserved \cite{I-cons, Gogberashvili:2016wsa, Gogberashvili:2021gfh, Gogberashvili:2022cam}.

Thus, in various scenarios, the Friedmann equation (\ref{balance}) contains the integration constant $C$ (the dark energy term). This change of the role of $\Lambda$ from a parameter of the matter action to a property of states does not solve the cosmological constant problem, but it does change it from a question of fine-tuning to a question of boundary conditions \cite{Buchmuller:2022msj, Gogberashvili:2016llo}. To find proper boundary conditions, note that cosmological equations are written for the Hubble volume, since as a proper causal boundary of the classical space-time (for the flat universe, $k = 0$) usually the Hubble horizon is considered \cite{FRW-bound-1, FRW-bound-2}. Then the metric fluctuations are bounded by $R_H$ and thermodynamic laws also are satisfied on this boundary \cite{Termo-bound-1, Termo-bound-2}.

On the other hand, the quantum fluctuations of matter fields should be limited not by the Hubble horizon $R_H$, but by the event horizon $R_e \ge R_H$, which represents a real boundary of space-time \cite{Gaztanaga:2021bgb, Gaztanaga:2022ktb}. Then, cosmological equations should also contain the energy density corresponding to entanglements of quantum particles across the Hubble horizon \cite{Mukohyama:1996yi} and can be taken into account by introducing a surface term at $R_H$. It was found that the perfect fluid of entanglement has a negative pressure \cite{Lee:2007zq} and can be interpreted as the origin of dark energy \cite{Gogberashvili:2018jkg}.

The value of the event horizon at the current cosmic time can be estimated as (see, for example, \cite{Margalef-Bentabol:2012kwa}):
\begin{equation} \label{R_e}
R_e = \frac {1}{H_0}\int_{-1}^0 \frac {dy}{\sqrt{\Omega_m (1 + y)^3 + \Omega_\Lambda}} \approx \frac {0.96\, R_H}{\sqrt {\Omega_\Lambda}} ~,
\end{equation}
where $\Omega_m$ denotes the matter density. While particle entanglements can be effective up to $R_e $, in the context of cosmology (as well as in the context of black holes), (\ref{R_e}) is always defined globally (see more discussions in \cite{Li:2004rb}) and at the event horizon we can assume absence of matter,
\begin{equation} \label{Omega_m}
\Omega_m |_{R \to R_e} \to 0~.
\end{equation}
Using this assumption, we can fix the integration constant in (\ref{balance}):
\begin{equation} \label{C}
C = H^2|_{R \to R_e}  = \frac {1}{R_e^2} ~.
\end{equation}
Therefore, for the dark energy density we obtain the value
\begin{equation} \label{DE}
\frac {C}{H_0^2} = \frac {R_H^2}{R_e^2} \approx 1.08 \, \Omega_\Lambda ~,
\end{equation}
which almost coincides with the observed density of the dark energy (\ref{Observ}), within the uncertainties in measurements of $\Omega_m$ and $\Omega_\Lambda$ in (\ref{R_e}).

To summarize, in this paper we have noted that within the unimodular or thermodynamic approaches to gravity, in the Friedmann equation (\ref{balance}) the cosmological term appears as an integration constant, i.e. is not associated with the large vacuum expectation values and can be fixed from boundary conditions. Since the real boundary of space-time is the event horizon and not the Hubble sphere, we assume that cosmological equations should contain the terms corresponding to the entanglements of quantum particles across the apparent horizon. Using the fact that in a region enclosed by an event horizon, like a black hole, interior matter density should tend to zero at the horizon (\ref{Omega_m}), in the Friedmann equation (\ref{balance}) we fix the integration constant (\ref{C}) and obtain the value for the dark energy density (\ref{DE}) that almost coincides with the observations (\ref{Observ}).

%%%%%%%%%%%%%%%%%%%%%%%%%%%%%%%%%%%%%%%%%%%%%%%%%%%%%%%%%%%%%%%%%%%%%%%%%%%%%%%

\end{document}